\documentclass[aps,reprint]{revtex4-1}
\usepackage{latexsym}
\usepackage{amssymb}
\usepackage{dcolumn}
\usepackage{bm}
\usepackage{comment}
\usepackage[dvips]{graphicx}
\usepackage{amsmath}
\usepackage{mathrsfs}
\usepackage[switch,columnwise]{lineno}
\newcommand{\be}{\begin{equation}}
\newcommand{\ee}{\end{equation}}
\newcommand{\bea}{\begin{eqnarray}}
\newcommand{\eea}{\end{eqnarray}}
\newcommand{\bref}[1]{(\ref{#1})}

\newcommand{\la}{\langle}
\newcommand{\ra}{\rangle}

\begin{document}
\preprint{KEK-TH-2145}
\title{Hunting for $T$-violation and Majoranality of Neutrinos in Muon Decays}
\author{Takeshi Fukuyama}
\email{fukuyama@se.ritsumei.ac.jp}
\affiliation{Research Center for Nuclear Physics (RCNP), Osaka University, Ibaraki, Osaka, 567-0047, Japan}
\author{Sohtaro Kanda}
\email{sohtaro.kanda@riken.jp}
\affiliation{RIKEN, 2-1 Hirosawa, Wako, Saitama 351-0198, Japan}
\author{Daisuke Nomura}
\email{dnomura@post.kek.jp}
\author{Koichiro Shimomura}
\email{ksimomu@post.kek.jp}
\affiliation{KEK, 1-1 Oho, Tsukuba, Ibaraki 305-0801, Japan}
\date{\today}
\begin{abstract}
We propose a new experiment to search for a time-reversal $(T)$ symmetry breaking process in muon decay and the Majoranality of the neutrinos.  In the presence of $V+A$ interactions, the Majoranality appears as a $T$-violating term in the muon decay width as shown by Doi et al \cite{Doi1982}, while in the Standard Model such a $T$-violating term is negligibly small.  The presences of $V+A$ interactions and the corresponding heavy right-handed Majorana neutrinos give us an important clue to solve two major issues in particle physics, the deficit of baryon asymmetry in the universe and the Majoranality of neutrinos. 
In the experiment, the polarization of positrons from muon decays is measured using a polarimeter consisting of a magnetized foil and a segmented calorimeter. According to our result of numerical calculation, a factor of ten improvement in sensitivity to the $T$-violating process is expected by a year of measurement at J-PARC Materials and Life Science Experimental Facility, compared to the most recent precursor experiment. 
\end{abstract}

\maketitle

\section{Introduction}
Just after the discovery of parity violation \cite{Lee1952, Wu1957}, possible tests of time reversal invariance ($T$-invariance) have been proposed \cite{Jackson1957}. Violation of $T$-invariance immediately implies $CP$-violation if invariance under $CPT$ transformation is assumed. $CP$-violation was afterward first found in kaon decays \cite{Christenson1964} and later in B-meson systems \cite{Belle2001, Babar2001}. These phenomena are successfully explained by the Kobayashi-Maskawa phase in the Cabibbo-Kobayashi-Maskawa (CKM) mixing matrix \cite{Kobayashi1973, Cabibbo1963}. Nevertheless, the observed baryon asymmetry in the universe (BAU) strongly indicates the existence of additional $CP$-violating factors beyond that of the Standard Model (SM).  

Also, in the lepton sector, we have recently obtained great achievements in understanding the generation structure,
including the determination of the Maki-Nakagawa-Sakata (MNS) mixing matrix elements \cite{Maki1962, Pontecorvo1957}. In these situations, it is one of the most important unsolved issues in particle physics whether the neutrinos are Majorana or Dirac particles. 

The most attractive idea to explain the tiny masses of the neutrinos is the seesaw mechanism via superheavy right-handed Majorana neutrinos \cite{Minkowski1977, Yanagida1979, GellMann1980, Glashow1980, Mohapatra1980}, indicating that the light neutrinos are Majorana particles. These superheavy neutrinos may have induced the observed baryon asymmetry via leptogenesis \cite{Fukugita1986}. In this scenario, decays of the superheavy neutrinos induce asymmetry in the lepton number, which is later converted to the baryon number asymmetry when the temperature of the universe is ${\cal O}(10) \text{TeV}$ due to the thermal fluctuation between vacua with different $B+L$ numbers \cite{Kuzmin1985}.
Thus, the superheavy right-handed neutrinos may solve the two major unsolved issues (the BAU and the neutrino Majoranality) mentioned above simultaneously. In this scenario, the existence of the right-handed weak current $V+A$ at very high energy is naturally expected since the right-handed neutrinos are most naturally predicted in many SO(10) grand unified theories (GUTs), where symmetry between the $V+A$ and $V-A$ currents are automatically predicted at high enough energies. This is also a robust prediction from many other GUTs.

In the presence of $V+A$ interactions, muon decays have a special implication since two emitted neutrinos interfere differently depending on whether they are Majorana or Dirac particles. Fortunately, the interference effect appears most clearly in the $CP$-violating term essentially without contamination of the SM background \cite{Doi1982}. Thus the search for $T$-violation in muon decay from both theoretical and experimental sides is a very attractive target to solve the above-mentioned problems, and is the motivation of this paper.

This paper is organized as follows. In Section \ref{section:section2}, we estimate the size of the $T$-violating term in the presence of $V+A$ current. It gives the relationship between $T$-violation and the Majorana property of the neutrinos. In Section \ref{section:section3} we discuss the $\nu$SM (SM+massive neutrinos) background on the $T$-violating term which appears from radiative corrections. This is indispensable in searching for effects from new physics beyond the SM (BSM physics). In the subsequent sections, we give an experimental proposal to detect the $T$-symmetry breaking term in muon decay and the Majoranality of the neutrinos.
 In Section \ref{section:section4}, a measurement principle to detect the transverse polarization of the positrons emitted from muon decays is described. In Section \ref{section:section5}, we study practical methods and feasibility of the proposed experiment. Section \ref{section:section6} is devoted to discussions, where the relation with another typical Majoranality experiment, neutrinoless double beta decay, is also argued. Section \ref{section:section7} is the summary.

\section{T-violating muon decays in New Physics beyond the SM}
\label{section:section2}
Before discussing muon decays in the presence of the $V+A$ current together with the usual $V-A$ one, we discuss the $V+A$ current in GUTs. As a BSM physics, GUTs are most promising and give comprehensive pictures. Among them, SO(10) GUT \cite{Fukuyama2013} is general enough. In the process of the symmetry breaking from SO(10) GUT to the SM, the right-handed current naturally appears in a symmetry breaking sequence like \footnote{The symmetry breaking to naive $\text{SU}(5)\otimes \text{U}(1)_{\text X}$ leads to too fast proton decay.}
\begin{align}
\rm SO(10)\to SU(4)_C\otimes SU(2)_L\otimes SU(2)_R\to \nonumber \\
\rm \text SU(3)_C\otimes SU(2)_L\otimes SU(2)_R\otimes U(1)_{B-L}\to SM.
\end{align}
This sequence shows that as the energy scale goes down, right-handed gauge bosons become massive at higher energy scale than the electroweak scale ($v_{\text{EW}}$) and the weak (left-handed) gauge bosons do so at $v_{\text{EW}}$. That is, left- and right-handed chiral Lagrangians successively appear as the energy scale goes up.

We should note that it is not clear which stage in the sequence above is directly related to the string scale where the gravitation is unified. For example, if we consider $ \text{SU}(4)_{\text C}\otimes \text{SU}(2)_{\text L}\otimes \text{SU}(2)_{\text R}\otimes \text{D}$ (Pati-Salam (PS) phase 
\cite{Pati1974} with D parity which interchanges left (L)- and right (R)-handed currents) as a remnant from SO(10), their coupling constants $g_{\text L},\ g_{\text R}$ and mixing matrices $U_{lj},\ V_{lj}$ satisfy following equations at the mass scale of GUT ($M_{\text{GUT}}$)
\begin{align}
g_{\text L}=g_{\text R}, ~~U_{lj}=V_{lj}.
\label{PS}
\end{align}
The mixing matrices $U_{lj}$ and $V_{lj}$ connect the weak $(l)$ and mass $(j)$ eigenstates. The former and the latter mixing matrices correspond to the left- and right-handed sectors, respectively. The constraints in Eq.~\bref{PS} are realized at $M_{\text{GUT}}$ but start to be violated as the energy goes down to the SM scale by renormalization group equations.

However, if the PS phase is the stage which is directly related to the string scale, we are free from the above constraints. Furthermore, if we start from $\text{SU}(3)_{\text C}\otimes \text{SU}(2)_{\text L}\otimes \text{SU}(2)_{\text R}\otimes \text{U}(1)_{\rm{B-L}}$, the quarks and leptons are separated and there are no leptoquark-type gauge bosons which connect the quarks and leptons.
If we forget about GUTs but still consider the
same mixing matrix as in Eq.~\bref{PS}, the
lower limit of $M_{\text{WR}}$ (mass of right-handed gauge boson) is relaxed to $M_{\text{WR}}>1.6$ TeV \cite{Beall1982}. 
Thus the right-handed currents appear in many BSM physics, where the $V+A$ interactions coexist with the $V-A$ interactions.  

Another theoretical focus of this paper is the Majoranality of neutrinos, which is the natural result of the seesaw mechanism of heavy right-handed neutrinos and Dirac neutrinos. Thus, as BSM physics, we consider the general effective interaction which includes the right-handed current
\begin{align}
H_{\text W}=\frac{G_{\text F}}{\sqrt{2}}\left[j_{{\text L}\alpha}^\dagger j_{\text L}^\alpha
   +\lambda j_{{\text R}\alpha}^\dagger j_{\text R}^\alpha
   +\kappa\left(j_{{\text L}\alpha}^\dagger j_{\text R}^\alpha
   +j_{{\text R}\alpha}^\dagger j_{\text L}^\alpha\right)\right],
\label{JLJR}
\end{align}
where $j_{{\text L}\alpha}$ and $j_{{\text R}\alpha}$ are given by
\begin{align}
j_{{\text L}\alpha} &\equiv 
 \sum_{l=e,\mu,\tau}\overline{l}(x)\gamma_\alpha(1-\gamma_5)
 \nu_{l{\text L}}(x),\\
j_{{\text R}\alpha} &\equiv \sum_{l=e,\mu,\tau}\overline{l}(x)\gamma_\alpha
   (1+\gamma_5)\nu_{l{\text R}}'(x).
\end{align}
Here $\nu_{\text L}~(\nu_{\text R}')$ are the left-handed (right-handed) weak eigenstates of neutrinos. The constants $\lambda$ and $\kappa$ are related to the mass eigenvalues of the weak bosons in the left- and right-handed gauge sector ($W_{\text L}, W_{\text R})$ as follows:
\begin{eqnarray} \label{LR1}
W_{\text L}&=&W_1\cos\zeta+W_2\sin\zeta,\\
W_{\text R}&=&-W_1\sin\zeta+W_2\cos\zeta,\\
\frac{G_{\text F}}{\sqrt{2}}&=&\frac{g^2}{8}\cos^2\zeta \frac{M_1^2\tan^2\zeta+M_2^2}{M_1^2M_2^2}\label{LR2},\\
\lambda &\equiv& \frac{M_1^2+M_2^2\tan^2 \zeta}{M_1^2\tan^2\zeta+M_2^2} \label{LR3},\\
\kappa &\equiv& \frac{-(M_2^2-M_1^2)\tan \zeta}{M_1^2\tan^2\zeta+M_2^2}\label{LR4},
\end{eqnarray}
where $M_1$ and $M_2$ are the masses of the mass eigenstates
$W_1$ and $W_2$, respectively, and $\zeta$ is the mixing angle
which relates the mass eigenstates and the gauge eigenstates.

\begin{figure}[h]
\centering
\includegraphics[width=\linewidth]{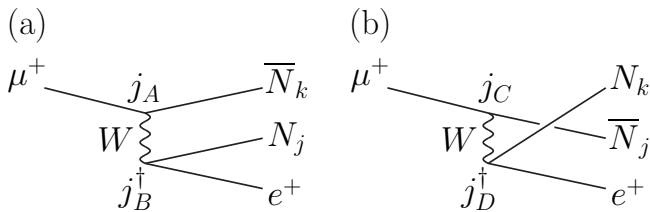}
\caption{
Tree level diagrams for muon decays.  In the diagrams, $A, \ldots, D$
denotes the handedness of the currents, $L$ or $R$.  In the set of
the diagrams with $(A,\ldots,D)=(L,L,R,R)$, the neutrinos with the
subscript $j$ in (a) and (b) have the same helicity, and those with the
subscript $k$ also have the same helicity, both in the massless limit.
The same statement also holds for the set with $(A,\ldots,D)=(R,R,L,L)$.
So the diagrams of each set interfere for the case of the Majorana
neutrinos.}
\label{fig:muondecay_intrfr}
\end{figure}

Using the effective Hamiltonian, the general form of the muon differential decay rate in the rest frame of $\mu^\mp$ is given by \cite{Doi1982}
\begin{align}
\frac{d\Gamma}{d{\bf q}_e}
&=\frac{m_\mu G_{\text F}^2}{3(2\pi)^4} \bigg\{ N(e)
 \pm\frac{ \left( \right. {\bf q}_e\cdot \boldsymbol{\zeta}_\mu \left. \right)}{E}P(e)
 \mp\frac{ \left( \right. {\bf q}_e\cdot \boldsymbol{\zeta}_e \left. \right)}{E}Q(e) \nonumber \\
&-\frac{({\bf q}_e\times \boldsymbol{\zeta}_\mu)\cdot
  \left( \right. {\bf q}_e\times \boldsymbol{\zeta}_e \left. \right)}{|{\bf q}_e |^2}R(e)
-\frac{ \left( \right. {\bf q}_e\cdot \boldsymbol{\zeta}_\mu \left. \right)
 \left( \right. {\bf q}_e\cdot \boldsymbol{\zeta}_e \left. \right)}{|{\bf q}_e |^2}S(e) \nonumber \\
&+ \frac{\boldsymbol{\zeta}_\mu\cdot
 \left( \right. {\bf q}_e\times\boldsymbol{\zeta}_e \left. \right)}{E}T(e) \bigg\},
\label{eq:drate}
\end{align}
where $E$ and ${\bf q}_e$ are the energy and momentum of the emitted electron (positron), and $\boldsymbol{\zeta}_e$ and $\boldsymbol{\zeta}_\mu$ are the polarization vectors of $e^{\mp}$ and $\mu^{\mp}$, respectively.  
In Eq.~(\ref{eq:drate}) we assume that we detect $e^{\mp}$ in the 
final state but not the neutrinos.
The upper and lower signs in Eq.~\bref{eq:drate} correspond to $\mu^-$ and $\mu^+$, respectively. This comes from the difference of the density matrix,
\begin{align}
\rho=\frac{1}{2}(\gamma^\mu q_\mu \pm m)(1+\gamma_5 \gamma^\nu a_\nu)~. 
\end{align}
Here the double-sign ``$\pm$'' corresponds to Eq.~\bref{eq:drate}. The four-dimensional spin vector $a_\mu$ is \cite{BLP}
\begin{align}
{\bf a}=\boldsymbol{\zeta}
+\frac{{\bf q}(\boldsymbol{\zeta}\cdot{\bf q})}{m(\epsilon+m)}, ~~~~
a^0=\frac{{\bf q}\cdot\boldsymbol{\zeta}}{m},
\end{align}
where $\boldsymbol{\zeta}$ is twice the mean spin vector in the rest frame and $\epsilon$ is the energy.
$N(e)$, $S(e)$, and $R(e)$ are $P$-even terms. $P(e)$ and $Q(e)$ are $P$-odd terms.  The $T(e)$ term is $P$-odd and $T$-odd ($i.e.$, $CP$-odd) term. Its explicit form is given by Eq.~\bref{Te}, whose presence indicates that the neutrinos are Majorana particles in the scheme of our $(V+A)$ $\oplus$ $(V-A)$ model. 
The dominant $\lambda$ term in Eq.~\bref{Te} comes from the interference of the 
diagrams of Fig.~\ref{fig:muondecay_intrfr} (a) and (b) 
with $(A,B,C,D)=(L,L,R,R)$ and $(R,R,L,L)$.  Here $L\equiv V-A$
and $R\equiv V+A$.  This interference takes place
only if the neutrinos are Majorana particles: Indeed, these diagrams indicate
the same reactions for Majorana neutrinos. 
The $T(e)$ term indicates BSM physics up to very tiny corrections due to the $\nu$SM. 

The explicit forms of $N(e),...,T(e)$ are given in \cite{Doi1982} and we give the only necessary term $T(e)$ in this paper, 
\begin{align}
T(e)=&-\epsilon_M\lambda h^I(m_\mu-m_e^2/m_\mu) \nonumber \\
&-3\kappa\left(v^Im_e-(v^I-4\kappa z^I)m_\mu\right) \nonumber \\
\approx &- \epsilon_M\lambda h^I(m_\mu-m_e^2/m_\mu),
\label{Te}
\end{align}
where $\epsilon_M=1$ and $\epsilon_M=0$ correspond to the cases that the neutrinos are Majorana and Dirac particles, respectively. The second equality comes from the fact that $v^I$ and $z^I$, whose explicit values are given in \cite{Doi1981}, are suppressed by ${m_i}/{m_\mu}$. Here, $m_i$ is the light neutrino mass, and are safely neglected except for around the maximum positron energy (see Eq.~(\ref{Emax})).
The parameter $h^I$ is defined as
\begin{align}
h^I\equiv \epsilon_M \sum_{j,k}F_{jk}^{1/2}G_{jk}\text{Im} C_{jk}'\approx \epsilon_M \sum_{j,k}\text{Im} C_{jk}',
\label{hI}
\end{align}
where
\begin{align}
\label{Fjk}
F_{jk}&=[1-(m_j+m_k)^2/\Delta^2][1-(m_j-m_k)^2/\Delta^2],\\
\label{Gjk}
G_{jk}&=1+(m_j^2+m_k^2)/\Delta^2-2(m_j^2-m_k^2)^2/\Delta^4,\\
\label{Cjk}
C'_{jk}&=U_{ej}^*V_{ek}V_{\mu j}^*U_{\mu k}
\end{align}
with 
\begin{align}
\Delta^2 \equiv 2m_\mu (W-E). 
\end{align}
Here $W$ is the maximum positron energy in the limit of $m_\nu=0$,
\begin{align}
W=\frac{m_\mu^2+m_e^2}{2m_\mu}.
\label{Emax}
\end{align} 
The imaginary part of $C_{jk}'$ is the ``generalized" Jarlskog invariant. The left and right-handed mixing matrices, $U$ and $V$, respectively, connect the weak eigenstates with their mass eigenstates as
\begin{align}
\nu_{l{\text L}}=\sum_{j=1}^{6}U_{lj}N_{j {\text L}},~~\nu_{l{\text R}}'=\sum_{j=1}^{6}V_{lj}N_{j {\text R}}.
\label{mixing}
\end{align} 
That is, the Jarlskog invariant of the CKM matrix \cite{Jarlskog1985} is replaced by that of the MNS mixing matrices of the left and right-handed neutrinos \cite{Maki1962}. This $T(e)$ term violates $T$-invariance. However, more importantly, only this term among $N(e),...,T(e)$ terms eventually vanishes if the neutrinos are not Majorana and is not contaminated by the $\nu$SM contributions. 
The magnitude of $\lambda$ is of order $(M_{\text{WL}}/M_{\text{WR}})^2$. The present lower limit of $M_{\text {WR}}$ is obtained from $\Delta L=2$ process \cite{Mandal2017} and of order $5$ TeV and $\lambda\leq 2.6\times 10^{-4}$ under the assumption of $V_{Rij}=\delta_{ij}$ \cite{PDG2016}. There still remains much larger possibility to detect the Majoranality than in the $\nu$SM case which will be shown in Eq.~\bref{SM}.

Here we add some comments on the $T$-violating term, $\boldsymbol{\zeta}_\mu\cdot({\bf q}_e\times \boldsymbol{\zeta}_e)$.  This term itself may appear as a result from new interactions other than $V+A$. For instance, if we generalize the Dirac bi-spinor coupling to $S$, $P$, $V$, $A$, $T$ (scalar, pseudo-scalar, vector, axial-vector, and tensor, respectively) using the Michel parameters \cite{Michel1950}, such a term appears even in the case of Dirac neutrinos. In our theory, a new physical entity is restricted in the heavy right-handed current and $T(e)$ term does not depend on the QCD vacuum angle. So the compatibility of the non-null result is immediately checked with other phenomena.

\begin{figure}[h]
\centering
\includegraphics[width=\linewidth]{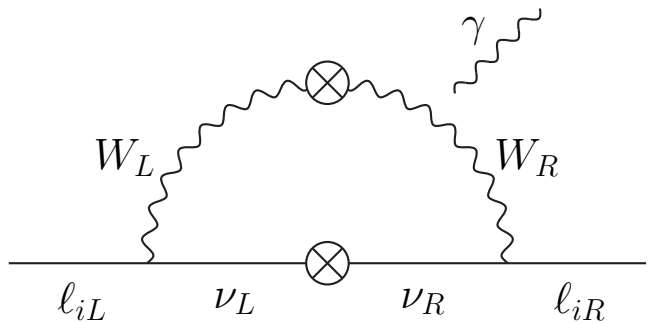}
\caption{The EDM diagram of the lepton $l_i$ $(i=1,\ldots,3)$ which appears at one-loop.}
\label{fig:LRedm}
\end{figure}

If there is the right-handed current, the lepton electric dipole moment (EDM) appears mainly from one loop diagram as shown in Fig.~\ref{fig:LRedm}, whereas it appears from four loop diagrams in the SM. 
The lepton EDM of Fig.~\ref{fig:LRedm} is estimated \cite{Bernreuther1991} with negligible generation mixing to be
\begin{align}
|d_{i}| <  8.2\times 10^{-27}\,
   \frac{|\text{Im}(m_{\text D})_{ii}|}{\text{MeV}}~ \text{e cm}, 
\end{align}
where $(m_{\text D})_{ii}$ denotes the $i$-th generation of the Dirac mass of the neutrino ($m_{\text D} \overline{\nu_{\text L}} \nu_{\text R}$). If $\text{SU}(3)_{\text C} \otimes \text{SU}(2)_{\text L} \otimes \text{SU}(2)_{\text R} \otimes \text{U}(1)_{\rm{B-L}}$ appears after the breaking of SO(10), $m_{\text D}$ is of order of up-type quark masses and the muon EDM may be significantly enhanced relative to the electron EDM.  The muon EDM is searched for at the muon $g-2$/EDM experiments using muon storage rings at Brookhaven National Laboratory (BNL) \cite{Bennett2006}, Fermi National Accelerator Laboratory (FNAL) \cite{Grange2015}, and Japan Proton Accelerator Research Complex (J-PARC) \cite{Abe2019}. 

\section{$T$-violating muon decays in Standard Model}
\label{section:section3}
To detect the signal discussed in the previous section, we have to clarify the main background processes to 
the signal process. Let us estimate the order of $T$-violating effect in the muon decay in the $\nu$SM framework.
\begin{figure}[htp]
\centering
\includegraphics[width=\linewidth]{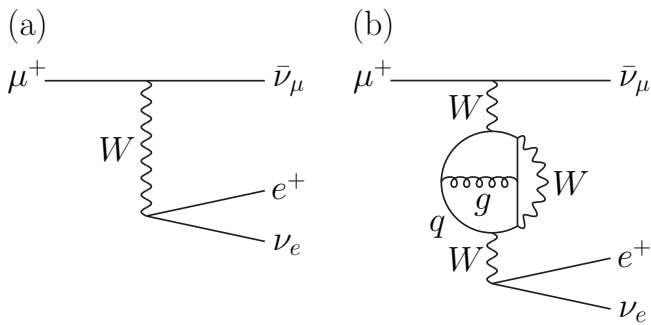}
\caption{(a) Tree level diagram of muon decays. 
(b) An example of the diagrams which violate $CP$. 
$CP$-violating term in muon decay rate appears as the interference term of (a) and (b).}
\label{fig:CPVSM}
\end{figure}
The main $T$-violating effect appears e.g.\ from the diagram (b) of Fig.~\ref{fig:CPVSM}, which interferes with the tree-level diagram, Fig.~\ref{fig:CPVSM} (a)~\footnote{
By using the same technique used in Ref.~\cite{Shabalin1978rs},
it can be shown that no $CP$ violation occurs if we neglect the gluon
in Fig.~\ref{fig:CPVSM} (b).}.  
This interference results in a $CP$-violating term in the muon differential decay width.

It is straightforward to estimate the size of the interference term between the two diagrams in Fig.~\ref{fig:CPVSM}. The $T$-violating decay width is suppressed by the factor of $\delta$ relative to the usual muon decay width, where
\begin{equation}
\delta=\frac{G_{\text F} m_t^2}{8\pi^2\sqrt{2}}
       \frac{\alpha_s}{4\pi}\frac{1}{16\pi^2}\frac{g^2}{8}J_{CP}
\approx 1\times 10^{-13}.
\label{SM}
\end{equation}
Here $J_{CP}$ is the Jarlskog parameter~\cite{Jarlskog1985}, which is a measure of $CP$-violation due to the CKM matrix~\cite{Kobayashi1973}, and $J_{CP} \approx 3.0 \times 10^{-5}$~\cite{PDG2016}.  We find that the $T$-violating effect in the muon decay in the $\nu$SM framework is extremely small.

\section{Experiment}
\label{section:section4}
We can search for a Majorana neutrino by measuring the transverse spin polarization of a positron from muon decay. In this section, an experimental principle and theoretical background of the experiment are described.

\subsection{Polarization of a positron from muon decay}
\begin{figure}
\includegraphics[width=\linewidth]{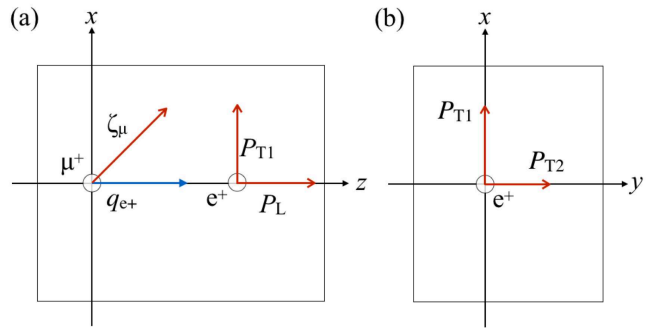}
\caption{Definition of the spin polarization of a positron from muon decay: (a) The muon spin $\boldsymbol {\zeta_\mu}$ and the momentum of decay positron ${\bf q}_e$ are on the $x$-$z$ plane; (b) On the transverse polarization, the $x$ and $y$ components are defined as $P_{\text T1}$ and $P_{\text T2}$, respectively. A non-zero $P_{\text T2}$ implies the $CP$ violation in muon decay.}
\label{fig:exp_pol}
\end{figure}

Figure \ref{fig:exp_pol} illustrates the definition of the $(x,y,z)$-coordinates in the muon rest frame. The positron polarization vector $\boldsymbol{\zeta}_e$ is described by three orthogonal components
\begin{eqnarray}
\boldsymbol{\zeta}_e=(P_{\text T1},P_{\text T2}, P_{\text L}) \equiv (\bm{P}_{\text T}, P_{\text L}),
\end{eqnarray}
where $P_{\text L}$ is the longitudinal component,  $P_{\text T1}$ and $P_{\text T2}$ are transverse components. The $z$-axis is defined in parallel to the positron momentum. In the SM, the positron polarization vector $\boldsymbol{\zeta}_e$ is coplanar with the positron momentum ${\bf q}_e$ and the muon spin $\boldsymbol{\zeta}_\mu$. In other words, only $P_{\text L}$ and $P_{\text T1}$ have non-vanishing norms. A non-zero $P_{\text T2}$ indicates the $CP$ violation in muon decay. 

\subsection{Polarimeter}
A precision polarimeter which utilizes the transverse spin-dependent interaction is required for the experiment. For positron polarimetry, five methods have been established: Mott scattering, Bhabha scattering, Compton scattering, positron annihilation-in-flight, and Compton transmission. Each technique utilizes the spin-dependent cross section of a particular electromagnetic process. 
The polarimeters are qualified by an analyzing power
\begin{align}
{\cal{A}} = \frac{N_+ - N_-}{N_+ + N_-},
\end{align}
where $N_+$ and $N_-$ are the number of positrons polarized in parallel and anti-parallel to the axis of interest, respectively. Statistical precision of the polarimetry is evaluated by the figure of merit (FOM) 
\begin{align}
\label{eq:FOM}
F = N {\cal A}^2,
\end{align}
where $N$ is the total number of positrons, $N=N_{+} + N_{-} $. Typical kinetic energy of incident positron ranges from 30 MeV to 50 MeV.

The analyzing power of polarimeter depends on positron energy.  In the energy region of interest, the polarimeters utilizing Bhabha scattering and annihilation-in-flight are feasible. The details of these polarimeters are described in following sub-sections.

\subsection{Bhabha scattering}
\begin{figure}
\includegraphics[width=\linewidth]{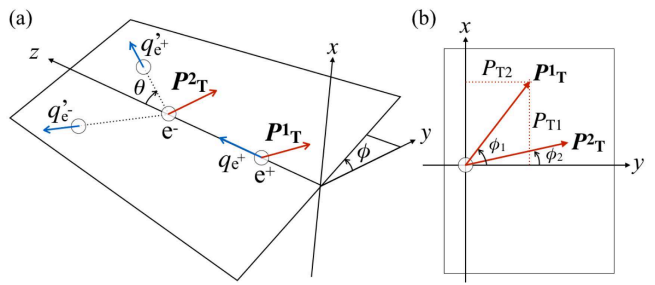}
\caption{Schematic of Bhabha scattering: (a) description of the scattering plane; (b) polarization vectors of an incident positron and a target electron. The $xyz$-frame is defined in Fig.~\ref{fig:exp_pol}.}
\label{fig:exp_scat}
\end{figure}

The Bhabha polarimeter is based on the spin-spin interaction between a positron spin and an electron spin. A thin foil of magnetized iron is commonly employed as a scattering target. It is sensitive to both transverse and longitudinal polarization components.  In the high energy limit, the differential cross-section of the Bhabha scattering in the center-of-mass system is \cite{Swartz1988}
\begin{align}
\label{Bhabha}
\left(\frac{d \sigma}{d \Omega}\right)_{\rm BHA} 
&= \frac{\alpha^2 (3+\cos^2 \theta^*)^2}{4s (1-\cos \theta^*)^2}
\left( \right. 1 - P_{\rm L}^{1} P_{\rm L}^{2} A_{\rm L}(\theta^*)  \nonumber \\
&- |\bm{P}_{\rm T}^{1}| | \bm{P}_{\rm T}^{2}| A_{\rm T}(\theta^*) 
   \cos(2 \phi - \phi_1 - \phi_2) \left. \right).
\end{align}
Here $\alpha$ is the fine structure constant, $\theta^*$ is the scattering angle in the center-of-mass frame, $\sqrt{s}$ is the center-of-mass energy,  and $\phi$ is the azimuth of the scattered positron. The superscripts $1$ and $2$ of the polarizations indicate the ones of beam and target, respectively. Figure \ref{fig:exp_scat} depicts the scattering plane. The azimuths $\phi_1$ and $\phi_2$ are the angles of the transverse polarizations $\bm{P}_{\rm T}^{1}$ and $\bm{P}_{\rm T}^{2}$ from the $y$-axis, respectively.

The longitudinal and transverse asymmetries $A_{\text L}$ and $A_{\text T}$ are determined as follows
\begin{align}
A_{\text L} = \frac{(7+ \cos^2 \theta^*)\sin^2 \theta^*}{(3+\cos^2 \theta^*)^2},
~~
A_{\text T} = \frac{\sin^4 \theta^*}{(3+\cos^2 \theta^*)^2}.
\end{align}
The analyzing power is energy independent and maximal at $\theta^*=90$ degrees in the center-of-mass system for both longitudinal and transverse polarizations.  The ratio of the transverse asymmetry to the longitudinal one depends on $\theta^*$ and $\phi$ for given $\phi_1$ and $\phi_2$ as
\begin{align}
 \frac{A_{\text T}(\theta^*) \cos (2 \phi - \phi_1 - \phi_2)}
      {A_{\text L}(\theta^*)} 
= \frac{\sin^2 \theta^*}{7 + \cos^2\theta^*}
  \cos(2 \phi - \phi_1 - \phi_2),
\label{eq:AT_AL_ratio}
\end{align}
which is the maximum at $(\theta^*, \phi)=(\pi/2, (\phi_1 + \phi_2)/2)$.

In the rest frame of the initial-state electron, which we call the lab frame for short, the scattering angle
$\theta_{\text{lab}}$ of the positron is related to $\theta^*$ by
\begin{align}
 \theta_{\text{lab}} = \arctan \left( \frac1{\gamma} \tan \frac{\theta^*}{2} \right),
\end{align}
where $\gamma \equiv \sqrt{s}/(2m_e)$.  The angle $\theta^*=\pi/2$ corresponds to  $\theta_{\text{lab}}=\arctan(2m_e/\sqrt{s})$.
It follows that, in the lab frame, both the analyzing power and the ratio Eq.~(\ref{eq:AT_AL_ratio})
take their maximum values at $(\theta_{\text{lab}}, \phi)= (\arctan(2m_e/\sqrt{s}), (\phi_1+\phi_2)/2)$.
Later in this paper, feasibility of a new experiment to detect the positron's transverse polarization will be discussed. 
There, the analyzing power and FOM of the polarimeters will be evaluated by using the GEANT4 software toolkit \cite{GEANT4_1, GEANT4_2, GEANT4_3}. Since interactions relevant to the longitudinal polarization are well established in GEANT4, firstly, we compute the longitudinal asymmetry for each small solid angle element $d\Omega = d(\cos \theta^*) d\phi$ and multiply the result by the factor $A_{\text T}(\theta^*) \cos (2 \phi - \phi_1 - \phi_2)/A_{\text L}(\theta^*)$ to calculate the transverse asymmetry.
In the lab frame, this conversion factor becomes largest for the very forward region, $(\theta_{\text{lab}}, \phi) \simeq  (\arctan(2m_e/\sqrt{s}), (\phi_1+\phi_2)/2)$.

\subsection{Annihilation-in-flight}
The annihilation-in-flight (AIF) polarimeter is based on the spin-dependent cross section of the positron annihilation. A scattering target is magnetized iron foil similarly to the case for the Bhabha polarimeter. The differential cross-section of AIF in the center-of-mass system in high energy limit is given by 
\begin{align}
\left(\frac{d\sigma}{d\Omega}\right)_{\rm AIF}
&=\frac{\alpha^2(1+\cos^2\theta^*)}
{s \sin^2\theta^*}\left( \right.
  1 + P_{\text L}^{1}P_{\text L}^{2}B_{\text L}(\theta^*) \nonumber \\
&+ |\bm{P}_{\text T}^{1}||\bm{P}_{\text T}^{2}|B_{\text T}(\theta^*)\cos (2\phi-\phi_1-\phi_2)
\left. \right),
\end{align}
where the definitions of $\theta^*$ and $\phi_i$ are same as those
in Eq.~(\ref{Bhabha}). Here $\phi$ is the azimuthal angle of one of the emitted photon.  Without loss of generality, we can take the range of $\phi$ as $-90<\phi/{\text {deg.}} \leq 90 $. The functions $B_{\text L}(\theta^*)$ and $B_{\text T}(\theta^*)$ are defined by the references \cite{Page1957, McMaster1961}
\begin{align}
B_{\text L}(\theta^*)=1, 
~~
B_{\text T}(\theta^*)=\frac{\sin^2\theta^*}{1+\cos ^2\theta^*}.
\end{align}
The AIF and Bhabha measurements can be performed simultaneously with the common apparatus. In contrast to the Bhabha polarimeter, the analyzing power of the AIF polarimeter for longitudinal and transverse polarizations are comparable. The analyzing power of the AIF polarimeter was calculated by Fetscher in detail \cite{Fetscher2007}. However, we emphasize that $P_{\text T2}$ appears even if his $F_{\text T2}$ vanishes because of the $V+A$ interaction and the Majoranality of the neutrinos.

Similarly to the case of Bhabha scattering, the ratio of the transverse asymmetry to the longitudinal one is given by
\begin{align}
 \frac{B_{\rm T}(\theta^*) \cos (2 \phi - \phi_1 - \phi_2)}
      {B_{\rm L}(\theta^*)} 
= \frac{\sin^2 \theta^*}{1 + \cos^2\theta^*}
  \cos(2 \phi - \phi_1 - \phi_2),
\label{eq:BT_BL_ratio}
\end{align}
which takes the largest value of unity at $(\theta^*, \phi)=(\pi/2, (\phi_1 + \phi_2)/2)$. The scattering angle $\theta_{\text{lab}}$ of the positron in the lab frame is related to $\theta^*$ by
\begin{align}
 \theta_{\text{lab}} 
= \arctan \left( \frac{\sin \theta^*}{\gamma(\beta + \cos \theta^*)} \right),
\end{align}
where $\gamma \equiv \sqrt{s}/(2m_e)$ and $\beta \equiv \sqrt{1-4m_e^2/s}$.  The angle $\theta^*=\pi/2$ corresponds to $\theta_{\text{lab}}=\arctan(1/(\beta\gamma))$, where the ratio Eq.~(\ref{eq:BT_BL_ratio}) becomes largest when $\phi=(\phi_1+\phi_2)/2$.  When $\sqrt{s} \gg m_e$, the ratio becomes largest in the very forward region, $(\theta_{\text{lab}}, \phi) \simeq (\arctan(2m_e/\sqrt{s}), (\phi_1+\phi_2)/2)$.

\subsection{Azimuthal angle dependence}
The transverse asymmetries in both Bhabha scattering and AIF have azimuthal dependence $\cos (2 \phi - \phi_1 - \phi_2)$ where $\phi$ is the scattering/emission azimuth, $\phi_1$ is the azimuth of positron polarization, and $\phi_2$ is the azimuth of target polarization. The azimuth $\phi_1$ is defined by the ratio between $P_{ \text T1}$ and $P_{ \text T2}$. The transverse polarizations $P_{ \text T1}$ and $P_{ \text T2}$ come from the $R(e)$ and $T(e)$ terms in Eq.~(\ref{eq:drate}), respectively. Only the latter violates the time-reversal symmetry. 

In typical left-right symmetric models, the magnitude of the ratio $|T(e)/N(e)|$ is typically expected to be on the order of $(M_{\text{WL}}/M_{\text{WR}})^2$. When we assume the right-handed weak gauge boson mass $M_{\text{WR}}$ ranges from 1 TeV to 10 TeV, $|T(e)/N(e)| = \mathcal{O}(10^{-4}) - \mathcal{O}(10^{-2})$. On the other hand, $|R(e)/N(e)|$ is expected to be on the order of $m_e/m_\mu = \mathcal{O}(10^{-2})$. These ratios give expected ranges of $\phi_1$ from 45 degrees to 135 degrees or from -135 degrees to -45 degrees.
To test the $T$-symmetry, the direction of target polarization will be in parallel or anti-parallel to the $y$-axis. In the setup of Figs.~\ref{fig:exp_pol} and \ref{fig:exp_scat}, it is most advantageous to make the direction of the target polarization be in parallel or antiparallel to the $y$-axis.

\section{Possibility of a new experiment at J-PARC}
\label{section:section5}
In this section, a new measurement of the transverse polarization of positrons from muon decays with the high-intensity pulsed muon beam at J-PARC is proposed. At Materials and Life Science Experimental Facility (MLF) Muon Science Establishment (MUSE), the world-highest intensity of pulsed muon beam has been established. The high-intensity beam is beneficial not only for the statistical precision but also a shortening of measurement time and suppression of the systematic uncertainty arising from temporal variations of the environment.

\subsection{Precursor experiments at PSI}
The most recent experimental search for the transverse polarization of muon decay positrons was performed at Paul Scherrer Institute (PSI) \cite{Danneberg2005}. This precursor experiment employed the AIF polarimeter and the continuous muon beam which is delivered at random timing. As an annihilation target, a magnetized metal foil with a thickness of 1 mm was placed between the muon stopping target and the photon detector. A honeycomb array calorimeter using $\rm{Bi_{12} Ge O_{20}}$ (BGO) crystals was employed for measuring photons. The target polarization was $P_e = 7.2\ \%$ and the beam polarization was $P_\mu = 82\pm 2\ \%$. The number of analyzed annihilation events was 36.4 million.
The results were
\begin{align}
P_{\text T1}  &= (6.3 \pm 7.7 \pm 3.4) \times 10^{-3}, \\
P_{\text T2}  &= (-3.7 \pm 7.7 \pm 3.4) \times 10^{-3},
\end{align}
where the first error is statistical uncertainty and the second is systematics. The systematic error was dominated by uncertainty in energy calibration for the calorimeter and positron energy loss before an annihilation which results in the fluctuation of the positron polarization.
Note that another experiment for a measurement of the Michel parameter $\xi '' $ in polarized muon decay was performed with the apparatus aforementioned \cite{Prieels2014}. The result was $\xi''=0.981\pm 0.045_{\text {stat}}\pm 0.003_{\text{syst}}$.
The parameter $\xi''$ is concerned with the positron's longitudinal polarization $P_{\text L}$, whereas our concern in the proposed experiment is the positron's transverse polarization $P_{\text T2}$.

\subsection{New experiment at J-PARC}
The new experiment proposed in this paper comprises three major components: high-intensity pulsed muon beam at J-PARC, an electron-polarized scattering/annihilation target, and segmented positron and photon detectors.

\subsubsection{Pulsed muon beam}
In contrast to the continuous beam at PSI, the pulsed muon beam at J-PARC has a periodic timing structure by nature. For a case of an experiment with pulsed beam, no muon trigger counter is required because arrival of the muon beam is synchronized to the accelerator repetition. The repetition cycle is 25 Hz and the beam has double pulse structure with the full-width at half-maximum of 100 ns. The two bunches have a timing interval of 600 ns. The expected muon beam intensity at J-PARC MLF MUSE H-Line is $1\times10^8\ \mu^+/\rm{s}$ \cite{Kawamura2018}. 

In the precursor experiments with the continuous beam, a simultaneous measurement of the Bhabha scattering and the AIF was impossible because of the difference in the data taking trigger configuration. Signatures of the Bhabha and AIF events are an electron-positron pair and photon pairs, respectively. In the new experiment, both the Bhabha and AIF measurements can be performed simultaneously.

\subsubsection{Decay positron polarimeter}

\begin{figure}
\includegraphics[width=\linewidth]{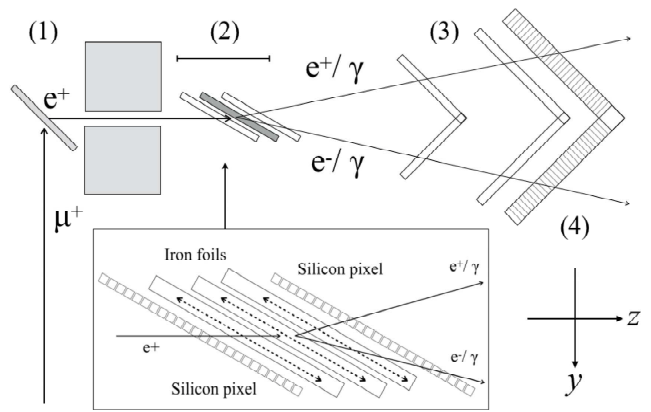}
\caption{Experimental schematic of the decay positron polarimeter: (1) the beryllium plate for muon stopping; (2) the spin-analyzing target between the silicon pixel detectors; (3) the silicon strip detectors for electron/positron tracking; (4) the segmented BGO calorimeter for gamma-ray detection. The inset in the bottom displays a magnified figure of the spin-analyzing target. The lead collimator is placed between the muon-stopping target and the spin-analyzing target. The dashed arrows indicate the magnetization directions of the iron foils. Note that the diagram is not drawn to scale.}
\label{fig:exp_setup}
\end{figure}
The polarimeter is designed for both the Bhabha and AIF measurements. Figure \ref{fig:exp_setup} depicts a schematic of the polarimeter. A thin plate of beryllium is placed for a muon-stopping target.  Magnetized iron foils are employed as a spin-analyzing target. Between the muon-stopping target and the spin-analyzing target, a lead collimator is placed to limit the incident angle of positrons.

In front and behind of the spin analyzing target, silicon pixel detectors are placed in order to determine the interaction point and the positron incident angle. An electron pair from Bhabha scattering and a gamma-ray pair from AIF are detected by silicon strip detectors and segmented BGO calorimeters, respectively. The muon-stopping target is placed at the angle of 45 degrees with respect to the muon beam direction for suppression of Coulomb multiple scattering inside the target. The spin-analyzing target is placed at the angle of 60 degrees because the magnetization of the iron foil is along to an in-plane direction. In the experiment, decay positrons are polarized in anti-parallel to the $z$-axis. The $z$-projection of the target polarization is in parallel or anti-parallel to the positron polarization. We denote the former and the latter cases as ``positive" and ``negative" configurations, respectively.

The measurement system is axial symmetric with respect to the direction of the incident muon beam except for the tilts of targets and the plurality of polarimeters can be placed to multiply statistics. A set of eight unit polarimeters under the symmetric arrangement is practical. Symmetric placement of the polarimeters is beneficial not only for statistics but also an understanding of systematic uncertainties arising from the target polarization and detector performances.

\subsection{Simulation study}
Feasibility of the designed polarimeter is evaluated by Monte-Carlo simulation using GEANT4. Performance of polarimeters is quantified by the figure of merit which is defined by Eq.~(\ref{eq:FOM}). In order to obtain the FOM, the number of signal events and angular asymmetries are calculated stepwise.

\subsubsection{Angular distributions}
The number of signal events per incident muon to the stopping target is calculated. For the case of Bhabha scattering, the signal events are selected by a pair of charged particle tracks, which is detected by the silicon tracker. For the case of AIF, the signals are discriminated by a pair of photons, which is detected by the calorimeters.

\begin{figure}
\includegraphics[width=0.9\linewidth]{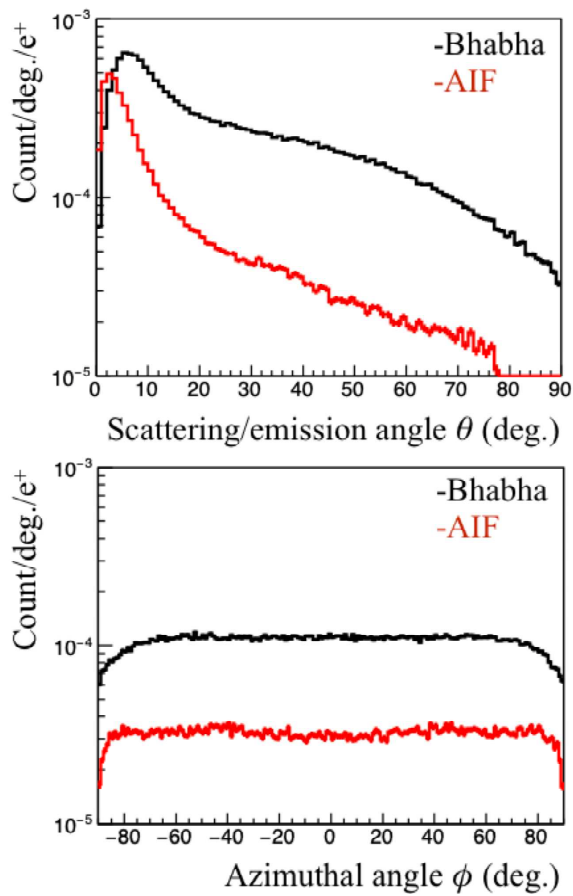}
\caption{Simulated angular distributions of Bhabha scattering and AIF: (top) scattering/emission angle ($\theta$); (bottom) azimuthal angle $\phi$. The histograms in black and red lines correspond to Bhabha scattering and AIF, respectively. The ordinates are normalized by the number of positrons on the spin-analyzing target. The ends of the distributions result from the boundary of simulated geometry.}
\label{fig:exp_theta}
\end{figure}

Simulated angular distributions are shown in Fig.~\ref{fig:exp_theta}. The result is obtained with the unpolarized spin-analyzing target with the thickness of 1 mm. Here and hereafter we denote the scattering/emission angle in the lab frame $\theta_{\text{lab}}$ as $\theta$ for short. In the simulation, positive muons having the kinetic energy of 4 MeV irradiate the muon-stopping target. The azimuthal angular distribution can be assumed uniform in the region of interest. 

\subsubsection{Angular asymmetries}
\begin{figure}
\includegraphics[width=0.9\linewidth]{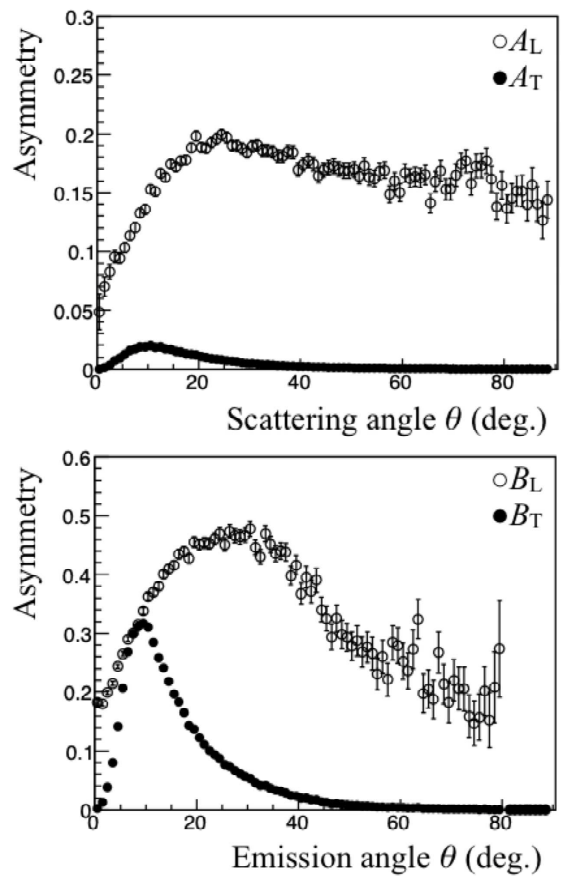}
\caption{Simulated asymmetries with Bhabha scattering and AIF: (top) the result of Bhabha polarimeter case; (bottom)  the result of AIF polarimeter case. The open and filled circles correspond to the longitudinal and transverse asymmetries, respectively.}
\label{fig:exp_asym}
\end{figure}

Spin-analyzing power of the polarimeter is numerically evaluated for polarized positrons. Figure \ref{fig:exp_asym} shows typical simulated asymmetries as functions of the scattering/emission angle $\theta$. The polarization of positrons from muon decay is assumed to be 100$\%$ in the longitudinal direction for calculation of the longitudinal asymmetry. The transverse asymmetry is calculated using the longitudinal asymmetry and the conversion factor aforementioned. Analyzing power for Bhabha scattering and AIF depend on the kinetic energies of electron-positron pair or two annihilation-quanta. Optima for the energy threshold for a measurement of the longitudinal polarization have been obtained to be 5.14 MeV and 6.66 MeV for the Bhabha polarimeter and AIF polarimeter, respectively \cite{Corriveau1981}. In our work, energy thresholds for a transverse polarization measurement are optimized with consideration of background processes.

Maximum transverse asymmetries are 0.02 for Bhabha scattering and 0.3 for AIF. This difference mainly resulted from background contamination and the factor of seven difference in the denominators in Eq.~(\ref{eq:AT_AL_ratio}) and Eq.~(\ref{eq:BT_BL_ratio}). Detection of a pair of an electron and a positron is likely to be affected by background processes such as bremsstrahlung rather than a pair of annihilation photons. Details of event selection will be described in the next subsection.
It is obvious that the AIF polarimeter is suitable for a measurement of the transverse polarization. However, the Bhabha polarimeter is still useful for measurements of the positron's longitudinal polarization and the muon beam polarization. Simultaneous measurements of the polarizations are important to understand the systematic uncertainties arising from beam quality and detector response. Hereafter, we will focus on the AIF polarimeter.

\subsubsection{Sensitivity to the $T$-violating effect}
In order to maximize the sensitivity for the transverse polarization, essential parameters in the experiment and analysis are optimized considering the FOM
\begin{equation}
\mathcal{F} = \frac{(N_+ - N_-)^2}{N_+ + N_-},
\end{equation}
where $N_+$ and $N_-$ are the numbers of observed events with ``positive" and ``negative" configurations in the target polarization, respectively. The thickness of the spin-analyzing target, the acceptances for the emission angle and the energy thresholds are optimized.

The number of signal events $N_+$ and $N_-$ are numerically calculated using the asymmetry $B_L$ shown in Fig.~\ref{fig:exp_asym} and Eq.~(\ref{eq:BT_BL_ratio}). The thickness of the spin-analyzing target, the energy thresholds for calorimeters, and the acceptance of the emission angle are optimized to maximize the FOM. In the optimization procedure, the thresholds are set for both the energies of individual photons ($E_1$ and $E_2$) and the sum of the photon energies ($E_1+E_2$).

\begin{figure}
\includegraphics[width=0.95\linewidth]{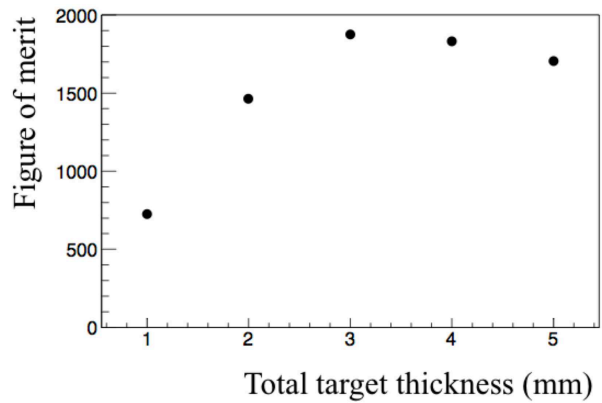}
\caption{Calculated target thickness dependence of the FOM. At each target thickness, the energy thresholds and the emission angle acceptance are optimized to maximize the FOM. The number of simulated muons is $1\times 10^7$ for each point.}
\label{fig:exp_tgt}
\end{figure}

Figure \ref{fig:exp_tgt} shows the target thickness dependence of the FOM. The target thickness of 3 mm is the optimum. For this case, the optimized energy thresholds are $E_1 > 2$ MeV, $E_2 > 2$ MeV, and $(E_1+E_2)>$ 15 MeV. The acceptance for $\theta$ is set $13\pm8$ degrees. Practically, the spin-analyzing target with the thickness of 3 mm will be consisting of three layers of 1 mm thick foils.

\begin{figure}
\includegraphics[width=0.95\linewidth]{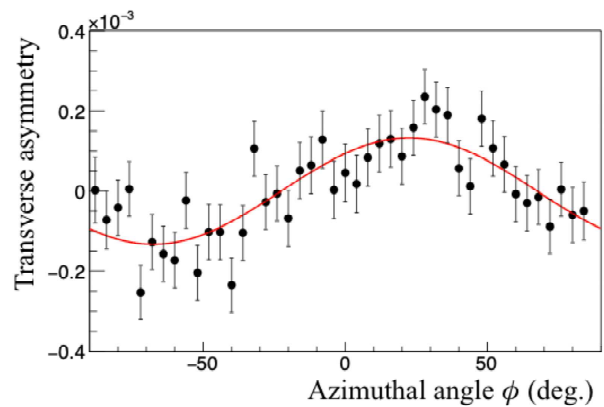}
\caption{Simulated transverse asymmetry as a function of the emission azimuth $\phi$. The red solid line indicates the fitting result by a cosine function. The oscillation amplitude obtained by the fitting is $(1.5 \pm 0.2 ) \times 10^{-4}$}
\label{fig:exp_cos}
\end{figure}

As described in Section \ref{section:section4}, the transverse asymmetry depends on the angles $\phi$, $\phi_1$, and $\phi_2$. Among these angles, the target polarization azimuth $\phi_2$ can be chosen at our disposal and the emission azimuth $\phi$ is obtained by the segmented calorimeter. Hence, we can extract $\phi_1$ from the observed azimuthal dependence of the transverse asymmetry. Figure \ref{fig:exp_cos} shows an example of expected azimuthal dependence for a case of $P_{\text T2} = 7.7 \times 10^{-3}$ that corresponds to statistical uncertainty in the precursor experimental result \cite{Danneberg2005}. The target polarization $\phi_2$ is set zero and $\phi_1$ is assumed to be 45 degrees. The number of simulated muons is $3.3 \times 10^9$, which corresponds to ten days of data acquisition at J-PARC.

We define the statistical significance of $P_{\text{T2}}$ by
\begin{equation}
\sigma \equiv \frac{P_{\text{T2}}}{\delta P_{\text{T2}}} ,
\label{eq:sigma_def}
\end{equation}
where $\delta P_{\text{T2}}$ is the statistical uncertainty
of $P_{\text{T2}}$.  Since $P_{\text{T2}}$ can be written as
\begin{equation}
P_{\text{T2}} = c a \cos(\phi_1),
\end{equation}
where $c$ is a constant and $a$ is the amplitude of the fitted cosine function
in the $\phi_1$ vs observed transverse asymmetry plane like
Fig.~\ref{fig:exp_cos}, Eq.~(\ref{eq:sigma_def}) can be written 
as
\begin{equation}
\sigma
 = \frac{1}{ \sqrt{ (\delta a/a)^2 + (\delta \phi_1 \tan \phi_1)^2}}.
\end{equation}
In the analysis, $\sigma \ge 3$ is required for a statistically significant
observation of the transverse polarization $P_{\text{T2}}$.  The transverse polarization has two components ($P_{\text{T1}}$ and $P_{\text{T2}}$), and $\tan\phi_1$ gives the ratio of them. A deviation of $\phi_1$ from 90 degrees together with a non-vanishing amplitude $a$ means a non-zero $P_{\text{T2}}$, $i.e.$, a $T$-violating effect in muon decay. 

\begin{figure}
\includegraphics[width=0.95\linewidth]{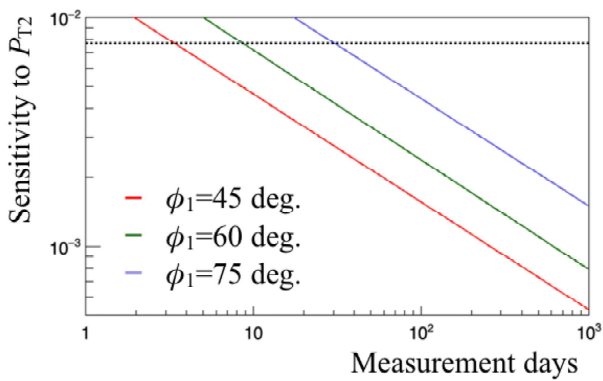}
\caption{Expected sensitivity to the transverse polarization using the AIF polarimeter. The abscissa represents the amount of time for $\sigma \ge 3$. The curves in each color correspond to the respective value of $\phi_1$. The horizontal dotted line indicates the statistical uncertainty for $P_{\text{T1}}$ and $P_{\text{T2}}$ in the precursor measurement \cite{Danneberg2005}.}
\label{fig:exp_sig}
\end{figure}

Amount of measurement time necessary for $\sigma \ge 3$ is estimated for several values of the transverse positron polarization $P_{\text{T2}}$. Figure \ref{fig:exp_sig} shows the sensitivity as a function of measurement time with the AIF polarimeter for $\phi_1=45, 60$ and 75 degrees.  When $|\phi_1-90^\circ|$ is large enough, the sensitivity is dominated by the requirement for the amplitude. On the other hand, the requirement for $\delta \phi_1$ is decisive for smaller $|\phi_1-90^\circ|$ cases. The muon beam intensity of $1\times10^8\ \mu^+/\rm{s}$ is assumed. The polarization of spin-analyzing target is set $7.2\%$. The target polarization direction is set ``positive" or ``negative" configurations and measurement time is equally divided into each configuration.
According to the estimation, when the magnitudes of $P_{\text{T1}}$ and $P_{\text{T2}}$ are similar, it is feasible to achieve a comparable statistical precision relative to that of the precursor experiment within four days.

\section{Discussions}
\label{section:section6}
The proposed experiment is feasible at the existing muon facility. However, further improvement is possible with the future facility upgrade. At J-PARC, construction of a new target station is being considered.  At the second target station, the muon beam intensity is expected to be several tens times higher than that at the present facility. The higher beam intensity contributes only to the statistical precision. The systematic uncertainty may be improved by using the magnetic resonance technique. In an analogy to the $\pi/2$-pulse method for nuclear magnetic resonance, the muon spin rotation will occur under a longitudinal static magnetic field and a pulse of radio-frequency field in the transverse direction \cite{Carne1984}. This technique enables us to rotate the muon spin polarization to an arbitrary direction. By combining the results of measurements with opposite rotation directions, the systematic uncertainty arising from the target polarization will be highly suppressed.

Lastly, we add some comments on an implication of $V+A$ interactions for neutrinoless double beta decay
($0\nu\beta\beta$) which is one of the most important experiments to detect the Majoranality of the neutrino \cite{Bilenky2010}.
In the absense of $V+A$ interactions,
it depends on the effective Majorana neutrino mass
\begin{align}
\la m_\nu\ra\equiv |\sum U_{ej}^2m_j|
\end{align}
in the $\nu$SM. In the presence of $V+A$ interactions, we have additionally $V+A$ effect and the inverse half-life time is given by \cite{Doi1985}
\begin{align}
\left(T^{0\nu}_{1/2}\right)^{-1}
&=|M^{0\nu}|^2 \biggl[ C_1\left(\frac{\la m_\nu\ra}{m_e}\right)^2
+C_2\la\lambda \ra\frac{\la m_\nu\ra}{m_e}\cos \psi_1 \nonumber\\
&~~+C_3\la \eta \ra \frac{\la m_\nu\ra}{m_e}\cos\psi_2+C_4\la\lambda\ra^2+C_5 \la \eta\ra^2 \nonumber\\
&~~+C_6\la \lambda\ra\la\eta\ra\cos (\psi_1-\psi_2) \biggr].
\label{eq:tmpEq42}
\end{align}
Here the explicit forms of $C_i$ and $M^{0\nu}$ are given in the reference \cite{Doi1985}. The parameters $\la\lambda\ra$ and $\la\eta\ra$ are defined by
\begin{align}
\la\lambda\ra=\lambda|\sum U_{ej}V_{ej}(g_{\text V}'/g_{\text V})|,
~~
\la\eta\ra=\eta|\sum U_{ej}V_{ej}|,
\end{align}
where the mixing matrices are defined in Eq.~\bref{mixing}. The parameters $g_{\text V}$ and $g_{\text V}'$ are the coupling constants of $V-A$ and $V+A$ hadronic currents, respectively.

The present lower limit of $T^{0\nu}_{1/2}$ is \cite{Gando2016}
\begin{equation}
T^{0\nu}_{1/2}>1.07\times 10^{26}\ \mbox{yr at 90\% C.L.}
\label{eq:halflife_limit}
\end{equation}
This implies
\begin{equation}
\la m_\nu\ra<61 \sim 165 ~\mbox{meV}
\end{equation}
for the case without $V+A$ interaction. 
In the presence of $V+A$, Eq.~(\ref{eq:halflife_limit})
gives a rather complex constraint from Eq.~(\ref{eq:tmpEq42})
and depends on the details of nuclear structures, whereas
our experiment is restricted to the leptonic world and the
measurement of $T(e)$ gives a rather simple relation on
Eq.~(\ref{Te}).   
Anyhow, $T$-violation is out of scope in $0\nu\beta\beta$ experiments.

\section{Summary}
\label{section:section7}
We have discussed the measurement of the $T$-violation in muon decay. The most important theoretical character is that it almost immediately indicates the Majoranality of neutrinos. This is based on the heavy right-handed neutrino, which is the necessary and robust object in most of the BSM physics models and also is the cause of the BAU via leptogenesis. Thus the $T$-violation (discovery of a non-zero $P_{\text T2}$) leads to the definitive information for the Majoranality of the neutrinos and BAU, two major unsolved problems in particle physics. We can check the distinction from the presence of the Michel parameters by estimating the order of BAU and comparing with other Majoranality experiments, as well as theoretical consistencies in many reliable models. 

The precursor experiment at PSI obtained the limits on the positron transverse polarizations with the sensitivity of $8 \times 10^{-3}$. We have proposed a new experiment with the high-intensity pulsed muon beam at J-PARC. The new experiment will employ the segmented annihilation-in-flight polarimeter with high-rate capability. The sensitivity and feasibility of the polarimeter have been studied numerically. We expect a factor of ten improvement in the sensitivity by a year of measurement.

\begin{acknowledgments}
We are greatly indebted to T. Mibe and H. Nishiura for useful discussions.
This work is supported by JSPS KAKENHI Grants Nos.17H01133 (K.S, T.F, and D.N) and 16K05323 (D.N).
\end{acknowledgments}

\bibliography{bib}

\begin{thebibliography}{49}%
\makeatletter
\providecommand \@ifxundefined [1]{%
 \@ifx{#1\undefined}
}%
\providecommand \@ifnum [1]{%
 \ifnum #1\expandafter \@firstoftwo
 \else \expandafter \@secondoftwo
 \fi
}%
\providecommand \@ifx [1]{%
 \ifx #1\expandafter \@firstoftwo
 \else \expandafter \@secondoftwo
 \fi
}%
\providecommand \natexlab [1]{#1}%
\providecommand \enquote  [1]{``#1''}%
\providecommand \bibnamefont  [1]{#1}%
\providecommand \bibfnamefont [1]{#1}%
\providecommand \citenamefont [1]{#1}%
\providecommand \href@noop [0]{\@secondoftwo}%
\providecommand \href [0]{\begingroup \@sanitize@url \@href}%
\providecommand \@href[1]{\@@startlink{#1}\@@href}%
\providecommand \@@href[1]{\endgroup#1\@@endlink}%
\providecommand \@sanitize@url [0]{\catcode `\\12\catcode `\$12\catcode
  `\&12\catcode `\#12\catcode `\^12\catcode `\_12\catcode `\%12\relax}%
\providecommand \@@startlink[1]{}%
\providecommand \@@endlink[0]{}%
\providecommand \url  [0]{\begingroup\@sanitize@url \@url }%
\providecommand \@url [1]{\endgroup\@href {#1}{\urlprefix }}%
\providecommand \urlprefix  [0]{URL }%
\providecommand \Eprint [0]{\href }%
\providecommand \doibase [0]{http://dx.doi.org/}%
\providecommand \selectlanguage [0]{\@gobble}%
\providecommand \bibinfo  [0]{\@secondoftwo}%
\providecommand \bibfield  [0]{\@secondoftwo}%
\providecommand \translation [1]{[#1]}%
\providecommand \BibitemOpen [0]{}%
\providecommand \bibitemStop [0]{}%
\providecommand \bibitemNoStop [0]{.\EOS\space}%
\providecommand \EOS [0]{\spacefactor3000\relax}%
\providecommand \BibitemShut  [1]{\csname bibitem#1\endcsname}%
\let\auto@bib@innerbib\@empty
\bibitem [{\citenamefont {Doi}\ \emph {et~al.}(1982)\citenamefont {Doi},
  \citenamefont {Kotani}, \citenamefont {Nishiura}, \citenamefont {Okuda},\
  and\ \citenamefont {Takasugi}}]{Doi1982}%
  \BibitemOpen
  \bibfield  {author} {\bibinfo {author} {\bibfnamefont {M.}~\bibnamefont
  {Doi}}, \bibinfo {author} {\bibfnamefont {T.}~\bibnamefont {Kotani}},
  \bibinfo {author} {\bibfnamefont {H.}~\bibnamefont {Nishiura}}, \bibinfo
  {author} {\bibfnamefont {K.}~\bibnamefont {Okuda}}, \ and\ \bibinfo {author}
  {\bibfnamefont {E.}~\bibnamefont {Takasugi}},\ }\href {\doibase
  10.1143/PTP.67.281} {\bibfield  {journal} {\bibinfo  {journal} {Progress of
  Theoretical Physics}\ }\textbf {\bibinfo {volume} {67}},\ \bibinfo {pages}
  {281} (\bibinfo {year} {1982})}\BibitemShut {NoStop}%
\bibitem [{\citenamefont {Lee}\ and\ \citenamefont {Yang}(1952)}]{Lee1952}%
  \BibitemOpen
  \bibfield  {author} {\bibinfo {author} {\bibfnamefont {T.~D.}\ \bibnamefont
  {Lee}}\ and\ \bibinfo {author} {\bibfnamefont {C.~N.}\ \bibnamefont {Yang}},\
  }\href {\doibase 10.1103/PhysRev.87.410} {\bibfield  {journal} {\bibinfo
  {journal} {Phys. Rev.}\ }\textbf {\bibinfo {volume} {87}},\ \bibinfo {pages}
  {410} (\bibinfo {year} {1952})}\BibitemShut {NoStop}%
\bibitem [{\citenamefont {Wu}\ \emph {et~al.}(1957)\citenamefont {Wu},
  \citenamefont {Ambler}, \citenamefont {Hayward}, \citenamefont {Hoppes},\
  and\ \citenamefont {Hudson}}]{Wu1957}%
  \BibitemOpen
  \bibfield  {author} {\bibinfo {author} {\bibfnamefont {C.~S.}\ \bibnamefont
  {Wu}}, \bibinfo {author} {\bibfnamefont {E.}~\bibnamefont {Ambler}}, \bibinfo
  {author} {\bibfnamefont {R.~W.}\ \bibnamefont {Hayward}}, \bibinfo {author}
  {\bibfnamefont {D.~D.}\ \bibnamefont {Hoppes}}, \ and\ \bibinfo {author}
  {\bibfnamefont {R.~P.}\ \bibnamefont {Hudson}},\ }\href {\doibase
  10.1103/PhysRev.105.1413} {\bibfield  {journal} {\bibinfo  {journal} {Phys.
  Rev.}\ }\textbf {\bibinfo {volume} {105}},\ \bibinfo {pages} {1413} (\bibinfo
  {year} {1957})}\BibitemShut {NoStop}%
\bibitem [{\citenamefont {Jackson}\ \emph {et~al.}(1957)\citenamefont
  {Jackson}, \citenamefont {Treiman},\ and\ \citenamefont
  {Wyld}}]{Jackson1957}%
  \BibitemOpen
  \bibfield  {author} {\bibinfo {author} {\bibfnamefont {J.~D.}\ \bibnamefont
  {Jackson}}, \bibinfo {author} {\bibfnamefont {S.~B.}\ \bibnamefont
  {Treiman}}, \ and\ \bibinfo {author} {\bibfnamefont {H.~W.}\ \bibnamefont
  {Wyld}},\ }\href {\doibase 10.1103/PhysRev.106.517} {\bibfield  {journal}
  {\bibinfo  {journal} {Phys. Rev.}\ }\textbf {\bibinfo {volume} {106}},\
  \bibinfo {pages} {517} (\bibinfo {year} {1957})}\BibitemShut {NoStop}%
\bibitem [{\citenamefont {Christenson}\ \emph {et~al.}(1964)\citenamefont
  {Christenson}, \citenamefont {Cronin}, \citenamefont {Fitch},\ and\
  \citenamefont {Turlay}}]{Christenson1964}%
  \BibitemOpen
  \bibfield  {author} {\bibinfo {author} {\bibfnamefont {J.~H.}\ \bibnamefont
  {Christenson}}, \bibinfo {author} {\bibfnamefont {J.~W.}\ \bibnamefont
  {Cronin}}, \bibinfo {author} {\bibfnamefont {V.~L.}\ \bibnamefont {Fitch}}, \
  and\ \bibinfo {author} {\bibfnamefont {R.}~\bibnamefont {Turlay}},\ }\href
  {\doibase 10.1103/PhysRevLett.13.138} {\bibfield  {journal} {\bibinfo
  {journal} {Phys. Rev. Lett.}\ }\textbf {\bibinfo {volume} {13}},\ \bibinfo
  {pages} {138} (\bibinfo {year} {1964})}\BibitemShut {NoStop}%
\bibitem [{\citenamefont {Abe}\ \emph {et~al.}(2001)\citenamefont {Abe} \emph
  {et~al.}}]{Belle2001}%
  \BibitemOpen
  \bibfield  {author} {\bibinfo {author} {\bibfnamefont {K.}~\bibnamefont
  {Abe}} \emph {et~al.} (\bibinfo {collaboration} {Belle Collaboration}),\
  }\href {\doibase 10.1103/PhysRevLett.87.091802} {\bibfield  {journal}
  {\bibinfo  {journal} {Phys. Rev. Lett.}\ }\textbf {\bibinfo {volume} {87}},\
  \bibinfo {pages} {091802} (\bibinfo {year} {2001})}\BibitemShut {NoStop}%
\bibitem [{\citenamefont {Aubert}\ \emph {et~al.}(2001)\citenamefont {Aubert}
  \emph {et~al.}}]{Babar2001}%
  \BibitemOpen
  \bibfield  {author} {\bibinfo {author} {\bibfnamefont {K.}~\bibnamefont
  {Aubert}} \emph {et~al.} (\bibinfo {collaboration} {BABAR Collaboration}),\
  }\href {\doibase 10.1103/PhysRevLett.87.091801} {\bibfield  {journal}
  {\bibinfo  {journal} {Phys. Rev. Lett.}\ }\textbf {\bibinfo {volume} {87}},\
  \bibinfo {pages} {091801} (\bibinfo {year} {2001})}\BibitemShut {NoStop}%
\bibitem [{\citenamefont {Kobayashi}\ and\ \citenamefont
  {Maskawa}(1973)}]{Kobayashi1973}%
  \BibitemOpen
  \bibfield  {author} {\bibinfo {author} {\bibfnamefont {M.}~\bibnamefont
  {Kobayashi}}\ and\ \bibinfo {author} {\bibfnamefont {T.}~\bibnamefont
  {Maskawa}},\ }\href {\doibase 10.1143/PTP.49.652} {\bibfield  {journal}
  {\bibinfo  {journal} {Progress of Theoretical Physics}\ }\textbf {\bibinfo
  {volume} {49}},\ \bibinfo {pages} {652} (\bibinfo {year} {1973})}\BibitemShut
  {NoStop}%
\bibitem [{\citenamefont {Cabibbo}(1963)}]{Cabibbo1963}%
  \BibitemOpen
  \bibfield  {author} {\bibinfo {author} {\bibfnamefont {N.}~\bibnamefont
  {Cabibbo}},\ }\href {\doibase 10.1103/PhysRevLett.10.531} {\bibfield
  {journal} {\bibinfo  {journal} {Phys. Rev. Lett.}\ }\textbf {\bibinfo
  {volume} {10}},\ \bibinfo {pages} {531} (\bibinfo {year} {1963})}\BibitemShut
  {NoStop}%
\bibitem [{\citenamefont {Maki}\ \emph {et~al.}(1962)\citenamefont {Maki},
  \citenamefont {Nakagawa},\ and\ \citenamefont {Sakata}}]{Maki1962}%
  \BibitemOpen
  \bibfield  {author} {\bibinfo {author} {\bibfnamefont {Z.}~\bibnamefont
  {Maki}}, \bibinfo {author} {\bibfnamefont {M.}~\bibnamefont {Nakagawa}}, \
  and\ \bibinfo {author} {\bibfnamefont {S.}~\bibnamefont {Sakata}},\ }\href
  {\doibase 10.1143/PTP.28.870} {\bibfield  {journal} {\bibinfo  {journal}
  {Progress of Theoretical Physics}\ }\textbf {\bibinfo {volume} {28}},\
  \bibinfo {pages} {870} (\bibinfo {year} {1962})}\BibitemShut {NoStop}%
\bibitem [{\citenamefont {Pontecorvo}(1958)}]{Pontecorvo1957}%
  \BibitemOpen
  \bibfield  {author} {\bibinfo {author} {\bibfnamefont {B.}~\bibnamefont
  {Pontecorvo}},\ }\href@noop {} {\bibfield  {journal} {\bibinfo  {journal}
  {Sov. Phys. JETP}\ }\textbf {\bibinfo {volume} {7}},\ \bibinfo {pages} {172}
  (\bibinfo {year} {1958})}\BibitemShut {NoStop}%
\bibitem [{\citenamefont {Minkowski}(1977)}]{Minkowski1977}%
  \BibitemOpen
  \bibfield  {author} {\bibinfo {author} {\bibfnamefont {P.}~\bibnamefont
  {Minkowski}},\ }\href {\doibase 10.1016/0370-2693(77)90435-X}
  {\bibfield  {journal} {\bibinfo  {journal} {Physics Letters B}\ }\textbf
  {\bibinfo {volume} {67}},\ \bibinfo {pages} {421 } (\bibinfo {year}
  {1977})}\BibitemShut {NoStop}%
\bibitem [{\citenamefont {Yanagida}(1979)}]{Yanagida1979}%
  \BibitemOpen
  \bibfield  {author} {\bibinfo {author} {\bibfnamefont {T.}~\bibnamefont
  {Yanagida}},\ }\href@noop {} {\bibfield  {journal} {\bibinfo  {journal}
  {Conf. Proc.}\ }\textbf {\bibinfo {volume} {C7902131}},\ \bibinfo {pages}
  {95} (\bibinfo {year} {1979})}\BibitemShut {NoStop}%
\bibitem [{\citenamefont {Gell-Mann}\ \emph {et~al.}(1979)\citenamefont
  {Gell-Mann}, \citenamefont {Ramond},\ and\ \citenamefont
  {Slansky}}]{GellMann1980}%
  \BibitemOpen
  \bibfield  {author} {\bibinfo {author} {\bibfnamefont {M.}~\bibnamefont
  {Gell-Mann}}, \bibinfo {author} {\bibfnamefont {P.}~\bibnamefont {Ramond}}, \
  and\ \bibinfo {author} {\bibfnamefont {R.}~\bibnamefont {Slansky}},\
  }\href@noop {} {\bibfield  {journal} {\bibinfo  {journal} {Conf. Proc.}\
  }\textbf {\bibinfo {volume} {C790927}},\ \bibinfo {pages} {315} (\bibinfo
  {year} {1979})}\BibitemShut {NoStop}%
\bibitem [{\citenamefont {Glashow}(1980)}]{Glashow1980}%
  \BibitemOpen
  \bibfield  {author} {\bibinfo {author} {\bibfnamefont {S.~L.}\ \bibnamefont
  {Glashow}},\ }\bibfield  {booktitle} {\emph {\bibinfo {booktitle} {Cargese
  Summer Institute: Quarks and Leptons Cargese, France, July 9-29, 1979}},\
  }\href@noop {} {\bibfield  {journal} {\bibinfo  {journal} {NATO Sci. Ser. B}\
  }\textbf {\bibinfo {volume} {61}},\ \bibinfo {pages} {687} (\bibinfo {year}
  {1980})}\BibitemShut {NoStop}%
\bibitem [{\citenamefont {Mohapatra}\ and\ \citenamefont
  {Senjanovi\ifmmode~\acute{c}\else \'{c}\fi{}}(1980)}]{Mohapatra1980}%
  \BibitemOpen
  \bibfield  {author} {\bibinfo {author} {\bibfnamefont {R.~N.}\ \bibnamefont
  {Mohapatra}}\ and\ \bibinfo {author} {\bibfnamefont {G.}~\bibnamefont
  {Senjanovi\ifmmode~\acute{c}\else \'{c}\fi{}}},\ }\href {\doibase
  10.1103/PhysRevLett.44.912} {\bibfield  {journal} {\bibinfo  {journal} {Phys.
  Rev. Lett.}\ }\textbf {\bibinfo {volume} {44}},\ \bibinfo {pages} {912}
  (\bibinfo {year} {1980})}\BibitemShut {NoStop}%
\bibitem [{\citenamefont {Fukugita}\ and\ \citenamefont
  {Yanagida}(1986)}]{Fukugita1986}%
  \BibitemOpen
  \bibfield  {author} {\bibinfo {author} {\bibfnamefont {M.}~\bibnamefont
  {Fukugita}}\ and\ \bibinfo {author} {\bibfnamefont {T.}~\bibnamefont
  {Yanagida}},\ }\href {\doibase 10.1016/0370-2693(86)91126-3}
  {\bibfield  {journal} {\bibinfo  {journal} {Physics Letters B}\ }\textbf
  {\bibinfo {volume} {174}},\ \bibinfo {pages} {45 } (\bibinfo {year}
  {1986})}\BibitemShut {NoStop}%
\bibitem [{\citenamefont {Kuzmin}\ \emph {et~al.}(1985)\citenamefont {Kuzmin},
  \citenamefont {Rubakov},\ and\ \citenamefont {Shaposhnikov}}]{Kuzmin1985}%
  \BibitemOpen
  \bibfield  {author} {\bibinfo {author} {\bibfnamefont {V.}~\bibnamefont
  {Kuzmin}}, \bibinfo {author} {\bibfnamefont {V.}~\bibnamefont {Rubakov}}, \
  and\ \bibinfo {author} {\bibfnamefont {M.}~\bibnamefont {Shaposhnikov}},\
  }\href@noop {} {\bibfield  {journal} {\bibinfo  {journal} {Physics Letters
  B}\ }\textbf {\bibinfo {volume} {155}},\ \bibinfo {pages} {36 } (\bibinfo
  {year} {1985})}\BibitemShut {NoStop}%
\bibitem [{\citenamefont {Fukuyama}(2013)}]{Fukuyama2013}%
  \BibitemOpen
  \bibfield  {author} {\bibinfo {author} {\bibfnamefont {T.}~\bibnamefont
  {Fukuyama}},\ }\href {https://doi.org/10.1142/S0217751X13300081} {\bibfield
  {journal} {\bibinfo  {journal} {Int.\ J.\ Mod.\ Phys.\ A}\
  }\textbf {\bibinfo {volume} {28}},\ \bibinfo {pages} {1330008} (\bibinfo
  {year} {2013})}\BibitemShut {NoStop}%
\bibitem [{Note1()}]{Note1}%
  \BibitemOpen
  \bibinfo {note} {The symmetry breaking to naive $\protect \text
  {SU}(5)\otimes \protect \text {U}(1)_{\protect \text X}$ leads to too fast
  proton decay.}\BibitemShut {Stop}%
\bibitem [{\citenamefont {Pati}\ and\ \citenamefont {Salam}(1974)}]{Pati1974}%
  \BibitemOpen
  \bibfield  {author} {\bibinfo {author} {\bibfnamefont {J.~C.}\ \bibnamefont
  {Pati}}\ and\ \bibinfo {author} {\bibfnamefont {A.}~\bibnamefont {Salam}},\
  }\href {\doibase 10.1103/PhysRevD.10.275} {\bibfield  {journal} {\bibinfo
  {journal} {Phys. Rev. D}\ }\textbf {\bibinfo {volume} {10}},\ \bibinfo
  {pages} {275} (\bibinfo {year} {1974})}\BibitemShut {NoStop}%
\bibitem [{\citenamefont {Beall}\ \emph {et~al.}(1982)\citenamefont {Beall},
  \citenamefont {Bander},\ and\ \citenamefont {Soni}}]{Beall1982}%
  \BibitemOpen
  \bibfield  {author} {\bibinfo {author} {\bibfnamefont {G.}~\bibnamefont
  {Beall}}, \bibinfo {author} {\bibfnamefont {M.}~\bibnamefont {Bander}}, \
  and\ \bibinfo {author} {\bibfnamefont {A.}~\bibnamefont {Soni}},\ }\href
  {\doibase 10.1103/PhysRevLett.48.848} {\bibfield  {journal} {\bibinfo
  {journal} {Phys. Rev. Lett.}\ }\textbf {\bibinfo {volume} {48}},\ \bibinfo
  {pages} {848} (\bibinfo {year} {1982})}\BibitemShut {NoStop}%
\bibitem [{\citenamefont {Berestetskii}\ \emph {et~al.}(1971)\citenamefont
  {Berestetskii}, \citenamefont {Lifshitz},\ and\ \citenamefont
  {Pitaevskii}}]{BLP}%
  \BibitemOpen
  \bibfield  {author} {\bibinfo {author} {\bibfnamefont {V.~B.}\ \bibnamefont
  {Berestetskii}}, \bibinfo {author} {\bibfnamefont {E.~M.}\ \bibnamefont
  {Lifshitz}}, \ and\ \bibinfo {author} {\bibfnamefont {L.~P.}\ \bibnamefont
  {Pitaevskii}},\ }\href@noop {} {\emph {\bibinfo {title} {Relativistic Quantum
  Theory}}}\ (\bibinfo  {publisher} {Pergamon Press},\ \bibinfo {year}
  {1971})\BibitemShut {NoStop}%
\bibitem [{\citenamefont {Doi}\ \emph {et~al.}(1981)\citenamefont {Doi},
  \citenamefont {Kotani}, \citenamefont {Nishiura}, \citenamefont {Okuda},\
  and\ \citenamefont {Takasugi}}]{Doi1981}%
  \BibitemOpen
  \bibfield  {author} {\bibinfo {author} {\bibfnamefont {M.}~\bibnamefont
  {Doi}}, \bibinfo {author} {\bibfnamefont {T.}~\bibnamefont {Kotani}},
  \bibinfo {author} {\bibfnamefont {H.}~\bibnamefont {Nishiura}}, \bibinfo
  {author} {\bibfnamefont {K.}~\bibnamefont {Okuda}}, \ and\ \bibinfo {author}
  {\bibfnamefont {E.}~\bibnamefont {Takasugi}},\ }\href@noop {} {\bibfield
  {journal} {\bibinfo  {journal} {Science Reports, Colleage of General
  Education, Osaka University}\ }\textbf {\bibinfo {volume} {30}},\ \bibinfo
  {pages} {119} (\bibinfo {year} {1981})}\BibitemShut {NoStop}%
\bibitem [{\citenamefont {Jarlskog}(1985)}]{Jarlskog1985}%
  \BibitemOpen
  \bibfield  {author} {\bibinfo {author} {\bibfnamefont {C.}~\bibnamefont
  {Jarlskog}},\ }\href {\doibase 10.1103/PhysRevLett.55.1039} {\bibfield
  {journal} {\bibinfo  {journal} {Phys. Rev. Lett.}\ }\textbf {\bibinfo
  {volume} {55}},\ \bibinfo {pages} {1039} (\bibinfo {year}
  {1985})}\BibitemShut {NoStop}%
\bibitem [{\citenamefont {Mandal}\ \emph {et~al.}(2017)\citenamefont {Mandal},
  \citenamefont {Mitra},\ and\ \citenamefont {Sinha}}]{Mandal2017}%
  \BibitemOpen
  \bibfield  {author} {\bibinfo {author} {\bibfnamefont {S.}~\bibnamefont
  {Mandal}}, \bibinfo {author} {\bibfnamefont {M.}~\bibnamefont {Mitra}}, \
  and\ \bibinfo {author} {\bibfnamefont {N.}~\bibnamefont {Sinha}},\ }\href
  {\doibase 10.1103/PhysRevD.96.035023} {\bibfield  {journal} {\bibinfo
  {journal} {Phys. Rev. D}\ }\textbf {\bibinfo {volume} {96}},\ \bibinfo
  {pages} {035023} (\bibinfo {year} {2017})}\BibitemShut {NoStop}%
\bibitem [{\citenamefont {Patrignani}\ and\ \citenamefont
  {Group}(2016)}]{PDG2016}%
  \BibitemOpen
  \bibfield  {author} {\bibinfo {author} {\bibfnamefont {C.}~\bibnamefont
  {Patrignani}}\ and\ \bibinfo {author} {\bibfnamefont {P.~D.}\ \bibnamefont
  {Group}},\ }\href {http://stacks.iop.org/1674-1137/40/i=10/a=100001}
  {\bibfield  {journal} {\bibinfo  {journal} {Chinese Physics C}\ }\textbf
  {\bibinfo {volume} {40}},\ \bibinfo {pages} {100001} (\bibinfo {year}
  {2016})}\BibitemShut {NoStop}%
\bibitem [{\citenamefont {Michel}(1950)}]{Michel1950}%
  \BibitemOpen
  \bibfield  {author} {\bibinfo {author} {\bibfnamefont {L.}~\bibnamefont
  {Michel}},\ }\href {http://stacks.iop.org/0370-1298/63/i=5/a=311} {\bibfield
  {journal} {\bibinfo  {journal} {Proc.\ Phys.\ Soc.\ A}\ }\textbf {\bibinfo {volume} {63}},\ \bibinfo {pages} {514} (\bibinfo
  {year} {1950})}\BibitemShut {NoStop}%
\bibitem [{\citenamefont {Bernreuther}\ and\ \citenamefont
  {Suzuki}(1991)}]{Bernreuther1991}%
  \BibitemOpen
  \bibfield  {author} {\bibinfo {author} {\bibfnamefont {W.}~\bibnamefont
  {Bernreuther}}\ and\ \bibinfo {author} {\bibfnamefont {M.}~\bibnamefont
  {Suzuki}},\ }\href {\doibase 10.1103/RevModPhys.63.313} {\bibfield  {journal}
  {\bibinfo  {journal} {Rev. Mod. Phys.}\ }\textbf {\bibinfo {volume} {63}},\
  \bibinfo {pages} {313} (\bibinfo {year} {1991})}\BibitemShut {NoStop}%
\bibitem [{\citenamefont {Bennett}\ \emph {et~al.}(2006)\citenamefont {Bennett}
  \emph {et~al.}}]{Bennett2006}%
  \BibitemOpen
  \bibfield  {author} {\bibinfo {author} {\bibfnamefont {G.~W.}\ \bibnamefont
  {Bennett}} \emph {et~al.} (\bibinfo {collaboration} {Muon g-2
  Collaboration}),\ }\href {\doibase 10.1103/PhysRevD.73.072003} {\bibfield
  {journal} {\bibinfo  {journal} {Phys. Rev. D}\ }\textbf {\bibinfo {volume}
  {73}},\ \bibinfo {pages} {072003} (\bibinfo {year} {2006})}\BibitemShut
  {NoStop}%
\bibitem [{\citenamefont {{Grange}}\ \emph {et~al.}(2015)\citenamefont
  {{Grange}} \emph {et~al.}}]{Grange2015}%
  \BibitemOpen
  \bibfield  {author} {\bibinfo {author} {\bibfnamefont {J.}~\bibnamefont
  {{Grange}}} \emph {et~al.},\ }\href@noop {} {\bibfield  {journal} {\bibinfo
  {journal} {arXiv:1501.06858}\ } (\bibinfo {year} {2015})}\BibitemShut
  {NoStop}%
\bibitem [{\citenamefont {{Abe}}\ \emph {et~al.}(2019)\citenamefont {{Abe}}
  \emph {et~al.}}]{Abe2019}%
  \BibitemOpen
  \bibfield  {author} {\bibinfo {author} {\bibfnamefont {M.}~\bibnamefont
  {{Abe}}} \emph {et~al.},\ }\href {\doibase 10.1093/ptep/ptz030} {\bibfield
  {journal} {\bibinfo  {journal} {PTEP}\ }\textbf {\bibinfo {volume} {2019}},\ \bibinfo {eid} {053C02}
  (\bibinfo {year} {2019})}\BibitemShut {NoStop}%
\bibitem [{Note2()}]{Note2}%
  \BibitemOpen
  \bibinfo {note} {By using the same technique used in Ref.~\cite
  {Shabalin1978rs}, it can be shown that no $CP$ violation occurs if we neglect
  the gluon in Fig.~\ref {fig:CPVSM} (b).}\BibitemShut {Stop}%
\bibitem [{\citenamefont {Swartz}(1988)}]{Swartz1988}%
  \BibitemOpen
  \bibfield  {author} {\bibinfo {author} {\bibfnamefont {M.}~\bibnamefont
  {Swartz}},\ }\href@noop {} {\bibfield  {journal} {\bibinfo  {journal} {SLAC
  Publication}\ }\textbf {\bibinfo {volume} {4656}} (\bibinfo {year}
  {1988})}\BibitemShut {NoStop}%
\bibitem [{\citenamefont {Agostinelli}\ \emph {et~al.}(2003)\citenamefont
  {Agostinelli} \emph {et~al.}}]{GEANT4_1}%
  \BibitemOpen
  \bibfield  {author} {\bibinfo {author} {\bibfnamefont {S.}~\bibnamefont
  {Agostinelli}} \emph {et~al.},\ }\href {\doibase
  10.1016/S0168-9002(03)01368-8} {\bibfield  {journal}
  {\bibinfo  {journal} {Nucl.\ Instrum.\ Meth.\ A}\ }\textbf {\bibinfo {volume} {506}},\ \bibinfo {pages} {250 }
  (\bibinfo {year} {2003})}\BibitemShut {NoStop}%
\bibitem [{\citenamefont {Allison}\ \emph {et~al.}(2006)\citenamefont {Allison}
  \emph {et~al.}}]{GEANT4_2}%
  \BibitemOpen
  \bibfield  {author} {\bibinfo {author} {\bibfnamefont {J.}~\bibnamefont
  {Allison}} \emph {et~al.},\ }\href {\doibase 10.1109/TNS.2006.869826}
  {\bibfield  {journal} {\bibinfo  {journal} {IEEE Trans.\ Nucl.\
  Sci.}\ }\textbf {\bibinfo {volume} {53}},\ \bibinfo {pages} {270}
  (\bibinfo {year} {2006})}\BibitemShut {NoStop}%
\bibitem [{\citenamefont {Allison}\ \emph {et~al.}(2016)\citenamefont {Allison}
  \emph {et~al.}}]{GEANT4_3}%
  \BibitemOpen
  \bibfield  {author} {\bibinfo {author} {\bibfnamefont {J.}~\bibnamefont
  {Allison}} \emph {et~al.},\ }\href {\doibase
  10.1016/j.nima.2016.06.125} {\bibfield  {journal} {\bibinfo
  {journal} {Nucl.\ Instrum.\ Meth.\ A}\
  }\textbf {\bibinfo {volume} {835}},\ \bibinfo {pages} {186 } (\bibinfo {year}
  {2016})}\BibitemShut {NoStop}%
\bibitem [{\citenamefont {Page}(1957)}]{Page1957}%
  \BibitemOpen
  \bibfield  {author} {\bibinfo {author} {\bibfnamefont {L.~A.}\ \bibnamefont
  {Page}},\ }\href {\doibase 10.1103/PhysRev.106.394} {\bibfield  {journal}
  {\bibinfo  {journal} {Phys. Rev.}\ }\textbf {\bibinfo {volume} {106}},\
  \bibinfo {pages} {394} (\bibinfo {year} {1957})}\BibitemShut {NoStop}%
\bibitem [{\citenamefont {McMaster}(1961)}]{McMaster1961}%
  \BibitemOpen
  \bibfield  {author} {\bibinfo {author} {\bibfnamefont {W.~H.}\ \bibnamefont
  {McMaster}},\ }\href {\doibase 10.1103/RevModPhys.33.8} {\bibfield  {journal}
  {\bibinfo  {journal} {Rev. Mod. Phys.}\ }\textbf {\bibinfo {volume} {33}},\
  \bibinfo {pages} {8} (\bibinfo {year} {1961})}\BibitemShut {NoStop}%
\bibitem [{\citenamefont {Fetscher}(2007)}]{Fetscher2007}%
  \BibitemOpen
  \bibfield  {author} {\bibinfo {author} {\bibfnamefont {W.}~\bibnamefont
  {Fetscher}},\ }\href {\doibase 10.1140/epjc/s10052-007-0384-6} {\bibfield
  {journal} {\bibinfo  {journal} {The European Physical Journal C}\ }\textbf
  {\bibinfo {volume} {52}},\ \bibinfo {pages} {1} (\bibinfo {year}
  {2007})}\BibitemShut {NoStop}%
\bibitem [{\citenamefont {Danneberg}\ \emph {et~al.}(2005)\citenamefont
  {Danneberg} \emph {et~al.}}]{Danneberg2005}%
  \BibitemOpen
  \bibfield  {author} {\bibinfo {author} {\bibfnamefont {N.}~\bibnamefont
  {Danneberg}} \emph {et~al.},\ }\href {\doibase 10.1103/PhysRevLett.94.021802}
  {\bibfield  {journal} {\bibinfo  {journal} {Phys. Rev. Lett.}\ }\textbf
  {\bibinfo {volume} {94}},\ \bibinfo {pages} {021802} (\bibinfo {year}
  {2005})}\BibitemShut {NoStop}%
\bibitem [{\citenamefont {Prieels}\ \emph {et~al.}(2014)\citenamefont {Prieels}
  \emph {et~al.}}]{Prieels2014}%
  \BibitemOpen
  \bibfield  {author} {\bibinfo {author} {\bibfnamefont {R.}~\bibnamefont
  {Prieels}} \emph {et~al.},\ }\href {\doibase 10.1103/PhysRevD.90.112003}
  {\bibfield  {journal} {\bibinfo  {journal} {Phys. Rev. D}\ }\textbf {\bibinfo
  {volume} {90}},\ \bibinfo {pages} {112003} (\bibinfo {year}
  {2014})}\BibitemShut {NoStop}%
\bibitem [{\citenamefont {Kawamura}\ \emph {et~al.}(2018)\citenamefont
  {Kawamura} \emph {et~al.}}]{Kawamura2018}%
  \BibitemOpen
  \bibfield  {author} {\bibinfo {author} {\bibfnamefont {N.}~\bibnamefont
  {Kawamura}} \emph {et~al.},\ }\href {\doibase 10.1093/ptep/pty116} {\bibfield
   {journal} {\bibinfo  {journal} {PTEP}\ }\textbf {\bibinfo {volume} {2018}},\ \bibinfo {pages} {113G01}
  (\bibinfo {year} {2018})}\BibitemShut {NoStop}%
\bibitem [{\citenamefont {Corriveau}\ \emph {et~al.}(1981)\citenamefont
  {Corriveau} \emph {et~al.}}]{Corriveau1981}%
  \BibitemOpen
  \bibfield  {author} {\bibinfo {author} {\bibfnamefont {F.}~\bibnamefont
  {Corriveau}} \emph {et~al.},\ }\href {\doibase 10.1103/PhysRevD.24.2004}
  {\bibfield  {journal} {\bibinfo  {journal} {Phys. Rev. D}\ }\textbf {\bibinfo
  {volume} {24}},\ \bibinfo {pages} {2004} (\bibinfo {year}
  {1981})}\BibitemShut {NoStop}%
\bibitem [{\citenamefont {Carne}\ \emph {et~al.}(1984)\citenamefont {Carne}
  \emph {et~al.}}]{Carne1984}%
  \BibitemOpen
  \bibfield  {author} {\bibinfo {author} {\bibfnamefont {A.}~\bibnamefont
  {Carne}} \emph {et~al.},\ }\href@noop {} {\bibfield  {journal} {\bibinfo
  {journal} {Hyperfine Interact}\ ,\ \bibinfo {pages} {17}} (\bibinfo {year}
  {1984})}\BibitemShut {NoStop}%
\bibitem [{\citenamefont {Bilenky}(2010)}]{Bilenky2010}%
  \BibitemOpen
  \bibfield  {author} {\bibinfo {author} {\bibfnamefont {S.~M.}\ \bibnamefont
  {Bilenky}},\ }\href {\doibase 10.1134/S1063779610050035} {\bibfield
  {journal} {\bibinfo  {journal} {Physics of Particles and Nuclei}\ }\textbf
  {\bibinfo {volume} {41}},\ \bibinfo {pages} {690} (\bibinfo {year}
  {2010})}\BibitemShut {NoStop}%
\bibitem [{\citenamefont {Doi}\ \emph {et~al.}(1985)\citenamefont {Doi},
  \citenamefont {Kotani},\ and\ \citenamefont {Takasugi}}]{Doi1985}%
  \BibitemOpen
  \bibfield  {author} {\bibinfo {author} {\bibfnamefont {M.}~\bibnamefont
  {Doi}}, \bibinfo {author} {\bibfnamefont {T.}~\bibnamefont {Kotani}}, \ and\
  \bibinfo {author} {\bibfnamefont {E.}~\bibnamefont {Takasugi}},\ }\href
  {\doibase 10.1143/PTPS.83.1} {\bibfield  {journal} {\bibinfo  {journal}
  {Progress of Theoretical Physics Supplement}\ }\textbf {\bibinfo {volume}
  {83}},\ \bibinfo {pages} {1} (\bibinfo {year} {1985})}\BibitemShut {NoStop}%
\bibitem [{\citenamefont {Gando}\ \emph {et~al.}(2016)\citenamefont {Gando}
  \emph {et~al.}}]{Gando2016}%
  \BibitemOpen
  \bibfield  {author} {\bibinfo {author} {\bibfnamefont {A.}~\bibnamefont
  {Gando}} \emph {et~al.} (\bibinfo {collaboration} {KamLAND-Zen
  Collaboration}),\ }\href {\doibase 10.1103/PhysRevLett.117.082503} {\bibfield
   {journal} {\bibinfo  {journal} {Phys. Rev. Lett.}\ }\textbf {\bibinfo
  {volume} {117}},\ \bibinfo {pages} {082503} (\bibinfo {year}
  {2016})}\BibitemShut {NoStop}%
\bibitem [{\citenamefont {Shabalin}(1978)}]{Shabalin1978rs}%
  \BibitemOpen
  \bibfield  {author} {\bibinfo {author} {\bibfnamefont {E.~P.}\ \bibnamefont
  {Shabalin}},\ }\href@noop {} {\bibfield  {journal} {\bibinfo  {journal} {Sov.
  J. Nucl. Phys.}\ }\textbf {\bibinfo {volume} {28}},\ \bibinfo {pages} {75}
  (\bibinfo {year} {1978})},\ \bibinfo {note} {[Yad.
  Fiz.28,151(1978)]}\BibitemShut {NoStop}%
\end{thebibliography}%
\end{document}